\documentclass[sigconf,nonacm]{acmart}
\renewcommand\footnotetextcopyrightpermission[1]{}
\AtBeginDocument{}

\usepackage{hyperref}
\usepackage{url}
\usepackage{booktabs}
\usepackage{amsfonts}
\usepackage{nicefrac}
\usepackage{microtype}
\usepackage[table]{xcolor}
\usepackage{amsmath}
\usepackage{graphicx}
\usepackage{capt-of}
\usepackage{multirow}
\usepackage{cleveref}
\usepackage{xspace}

\newcommand{\methodname}{ReST\xspace}

\definecolor{baselinegray}{gray}{0.96}

\title[Scaling Sequence Transformers for Recommendation with Rec-Native Designs]{From Language to Behavior: Scaling Sequence Transformers for Industrial Recommendation Ranking with Rec-Native Designs}

\author{Jie Chen, Xiangqian Yu, Yanchao Lian, Tan Lu, Run Yang, Zhengchun Shang,  \\ Xing Wang, Cheng Chen, Ke Hu,  Qiang Li\textsuperscript{*}, Tianjiu Yin, Xiaobing Liu}
\affiliation{\institution{ByteDance}
  \city{}
  \country{\{chenjie.996, yuxiangqian.0, lianyanchao, lutan, yangrun,austin.shang, xing.wang, chencheng.kit, xiaohe.00,\\huangtao.ranger, yintianjiu, will.liu\}@bytedance.com}
}

\renewcommand{\shortauthors}{Chen et al.}

\begin{document}

\begin{abstract}
Scaling Transformers has driven large gains in language modeling, but transplanting this to behavior-sequence modeling in production ranking is challenging: recommendation differs in signal quality, where behavior sequences are noisy, temporally irregular, and sparsely supervised, and in computation asymmetry, where each request scores many candidates against one shared user history under tight latency budgets. We propose ReST, a recommendation-native Transformer scaling framework. For signal quality, it introduces a sequence encoder with dual-gated attention, rotary positional and temporal embedding, stabilized residual normalization, and training-only auxiliary objectives. For computation asymmetry, it factorizes ranking into a heavy reusable encoder and a lightweight cross decoder with projection-free KV attention and token-specific parameterization, coupling user-level shared-prefix training with shared-prefix serving for compute-once, decode-many-times ranking. Across industrial and public benchmarks, ReST achieves higher accuracy and scales more consistently along sequence length, depth, and width, where LLM-style Transformer blocks saturate. A one-week online A/B test on a production advertising platform improves online AUC by 1.31\% and lifts a core revenue metric by 11.93\% within a 50\,ms P99 budget; ReST has since been fully deployed in production, showing that behavior-sequence scaling remains a promising, under-exploited axis for production ranking.
\end{abstract}




\maketitle
\begingroup
\renewcommand{\thefootnote}{\fnsymbol{footnote}}
\footnotetext[1]{Corresponding author.}
\endgroup

\section{Introduction}
\label{sec:intro}
Transformers have become the dominant scaling recipe in language modeling, where increasing model capacity, data, and training compute yields substantial and consistent gains~\cite{kaplan2020scaling, hoffmann2022chinchilla, llama2023}. Recommendation ranking appears to offer a similar opportunity, as user behavior sequences capture evolving intent beyond static user features. This promise has motivated a long line of behavior-sequence models, from target-aware attention~\cite{din2018} and long-sequence retrieval~\cite{sim2020, sdim2022} to recent large recommendation models~\cite{hstu2024}. However, directly scaling Transformer sequence models for production discriminative ranking remains challenging, because recommendation differs from language modeling in two fundamental ways.

The first obstacle lies in \emph{signal quality}. Unlike natural language, user behavior sequences exhibit three distinct characteristics that make naive scaling statistically inefficient. \emph{(i) Behavior reliability.} Behavior tokens are implicit feedback: clicks and views may reflect stable preference, but also accidental exposure, exploratory browsing, or biased feedback~\cite{joachims2017unbiased, wang2021denoising}. Standard self-attention aggregates them through a shared softmax without an explicit mechanism to distinguish reliable from unreliable behavioral evidence. \emph{(ii) Temporal structure.} User behaviors occur in irregular physical time---adjacent actions may be seconds or days apart, so equal ordinal distances imply very different recency---and purely ordinal encoding is insufficient for multi-scale recency~\cite{tisasrec2020}. \emph{(iii) Supervision density.} Decoder-only language models receive dense next-token supervision, whereas industrial ranking provides only sparse, delayed labels at the user and candidate level~\cite{chapelle2014delayed, ktena2019delayed}. This sparse discriminative supervision provides a less direct training signal to the sequence stack, and the difficulty is further amplified in hybrid rankers: the main discriminative loss can be shortcut through the strong non-sequential DLRM branch, leaving the sequence module weakly supervised even when the history carries useful intent---a \emph{sequence-starvation} effect that worsens as the module scales. Consequently, naively scaling a vanilla Transformer yields diminishing returns~\cite{guo2024scaling}.

The second obstacle lies in \emph{computation asymmetry} in target-conditioned ranking. Unlike language modeling, a ranking request pairs one user history with $N$ candidate targets~\cite{covington2016youtube, hstu2024}. The two sides benefit from scaling differently: sequence-side computation is shared across candidates, so investing more compute there can be amortized over all $N$ targets; target-conditioned interaction, however, depends on each candidate and is instantiated $N$ times per request, making per-candidate compute the latency bottleneck while still requiring enough capacity to discriminate fine-grained candidates. An LLM-style decoder-only Transformer places both roles in a single stack and couples computation with parameters, making it hard to serve these divergent needs separately. Scaling the sequence model under a production latency budget therefore requires breaking this symmetry.

We propose \textbf{\methodname}, a Recommendation-native Scalable Transformer framework that redesigns the Transformer scaling recipe for behavior-sequence modeling in production ranking.
To address the \emph{signal quality} challenge, \methodname introduces a recommendation-native sequence encoder block for behavior data, combining dual-gated attention, rotary positional and temporal embedding (RoPE + RoTE), stabilized residual normalization, and training-only auxiliary objectives that strengthen sequence-side supervision. To address \emph{computation asymmetry}, \methodname factorizes ranking into a compute-heavy sequence encoder \(T\), which contextualizes each user history once and produces reusable memory, and a lightweight cross decoder \(C\), which scores many candidates through projection-free key/value attention and token-specific query parameterization. We further co-design the training and serving stack around this boundary: user-level shared-prefix training amortizes prefix computation across samples from the same user, while shared-prefix serving reuses encoder states across candidates within a request. Empirically, \methodname restores consistent scaling along behavior sequence length, model depth, and hidden dimension, reduces redundant training and serving computation, and satisfies the production P99 latency budget in online deployment.

Our contributions are summarized as follows:

    \textbf{1. Rec-native Transformer for sequence scaling.}
    We identify and empirically diagnose a sequence-starvation effect in hybrid production rankers. We develop a recommendation-native sequence encoder that combines dual-gated attention for noisy behaviors, rotary positional and temporal embedding for irregular temporal structure, and SRN for stable depth scaling, together with two training-only auxiliary objectives that mitigate sequence starvation.

    \textbf{2. Asymmetric architecture with system co-design.}
    We exploit the one-history-to-many-candidates structure of production ranking by factorizing computation into a reusable, compute-heavy sequence encoder and a lightweight candidate-aware cross decoder. Projection-free key/value attention and token-specific parameterization preserve candidate-side capacity with low activated FLOPs, while user-level shared-prefix training and shared-prefix serving translate this architectural asymmetry into computation reuse.

    \textbf{3. Large-scale empirical validation and online deployment.}
    Across a large-scale industrial advertising dataset and public benchmarks, \methodname achieves better accuracy--efficiency trade-offs than LLM-style and recommendation baselines, and continues to scale in sequence length, depth, and width where LLM-style blocks saturate. A one-week online A/B test improves online AUC by 1.31\% and lifts a core revenue metric by $\mathbf{11.93\%}$ within the $\mathbf{50}$\,ms P99 latency budget, leading to full production deployment.

\section{Method}
\label{sec:method}

\subsection{Problem Formulation}
\label{subsec:problem}

We study conversion rate (CVR) prediction in large-scale advertising ranking, where industrial rankers combine chronological behavior sequences with non-sequential user/context features~\cite{din2018, dlrm2019}. Let $\mathcal{U}$ and $\mathcal{A}$ denote the user and ad sets. For a user $u \in \mathcal{U}$, we observe a behavior sequence $S_u = [(\mathbf{s}_1, t_1), \ldots, (\mathbf{s}_L, t_L)]$, where $\mathbf{s}_i$ is the feature vector of the $i$-th interaction and $t_i$ its physical timestamp, together with non-sequential features $\mathbf{u}_d$.

In production ranking, these shared user-side inputs are paired with $N$ candidate ads $\{a_1,\ldots,a_N\}\subseteq\mathcal{A}$. For each candidate $a$, the ranker predicts
\begin{equation}
\label{eq:overall}
    \hat{y} = f_{\theta}(S_u, \mathbf{u}_d, a) \approx P(y=1 \mid S_u, \mathbf{u}_d, a),
\end{equation}
where $y \in \{0,1\}$ is the binary conversion label. The model is trained with the CVR loss $\mathcal{L}_{\mathrm{ce}} = \mathbb{E}_{\mathcal{D}}\big[\ell_{\mathrm{BCE}}(\hat{y}, y)\big]$, where $\ell_{\mathrm{BCE}}$ is the standard binary cross-entropy over the training set $\mathcal{D}$.

This work focuses on scaling the sequence modeling component under this discriminative formulation, while keeping the non-sequential ranking stack and candidate scoring protocol intact.

\begin{figure*}[t]
    \centering
    \includegraphics[width=0.9\textwidth]{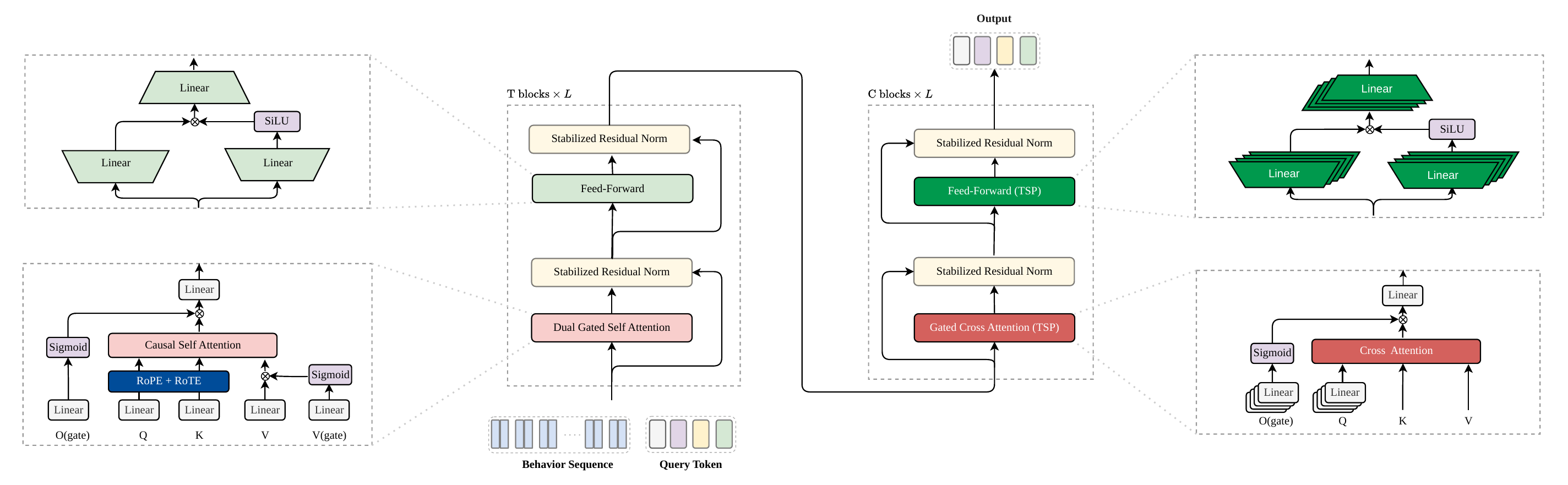}
    \caption{\methodname architecture: a compute-heavy sequence encoder \(T\) and a lightweight cross decoder \(C\).}
    \Description{Architecture diagram showing \methodname factorizing ranking into a reusable sequence encoder and a lightweight cross decoder.}
    \label{fig:rest}
\end{figure*}

\subsection{Overview}
\label{subsec:rest}

To address the signal quality and computation asymmetry challenges, we propose {\methodname}, a recommendation-native scalable Transformer for behavior-sequence modeling in recommendation ranking (Figure~\ref{fig:rest}). \methodname adopts an asymmetric encoder--decoder architecture: a compute-heavy rec-native sequence Transformer encoder $T$ builds robust, reusable memory once per user behavior history (Sec.~\ref{subsec:rec-native}), while a lightweight cross decoder $C$ performs candidate-specific decoding from this memory (Sec.~\ref{subsec:decoder}). We further introduce two training-only auxiliary objectives to mitigate sequence starvation (Sec.~\ref{subsec:aux-supervision}), together with user-level shared-prefix training and shared-prefix serving (Sec.~\ref{subsec:system}), enabling compute-once, decode-many-times ranking under production latency budgets.

\subsection{Rec-Native Sequence Encoder (\texorpdfstring{\(T\)}{T})}
\label{subsec:rec-native}
We start from a LLaMA-style~\cite{llama2023} Transformer block with causal self-attention and SwiGLU~\cite{shazeer2020glu} feed-forward layers, but adapt it to the distinct properties of recommendation behavior sequences. Naively scaling such blocks is hindered by noisy interactions, irregular temporal gaps, and unstable optimization as depth grows under sparse supervision. We therefore introduce three recommendation-native sequence modeling designs: Dual-Gated Attention for noisy behaviors, Rotary Positional and Temporal Embedding for ordinal and temporal structure, and Stabilized Residual Normalization for stable depth scaling.

\paragraph{Dual-Gated Attention (DGA)}
Implicit feedback in recommendation is inherently noisy~\cite{din2018, wang2021denoising}: accidental clicks, weak intent, and exploratory actions may not faithfully reflect stable user preference. This noise affects attention in two stages. Before aggregation, an individual behavior token may be unreliable evidence; after aggregation, the resulting behavioral context may still be only weakly useful for updating the representation. We therefore design Dual-Gated Attention (DGA) to make denoising explicit on both sides of the attention operator. A value gate performs pre-aggregation filtering over behavior features, while an output gate performs post-aggregation context modulation. Given an input sequence \(\mathbf{X} \in \mathbb{R}^{L \times d}\),
\begin{align}
    \mathbf{H}
    =
    \mathrm{Attention}\!\Bigl(
        \mathbf{X}\mathbf{W}_Q,\,
        \mathbf{X}\mathbf{W}_K,\,
        \sigma(\mathbf{X}\mathbf{W}_{gv}) \odot (\mathbf{X}\mathbf{W}_V)
    \Bigr),
    \\
    \mathrm{DualGatedAttn}(\mathbf{X})
    =
    \bigl(\mathbf{H} \odot \sigma(\mathbf{X}\mathbf{W}_{go})\bigr)\mathbf{W}_o,
\end{align}
where \(\mathrm{Attention}(\mathbf{Q},\mathbf{K},\mathbf{V})=\mathrm{softmax}(\mathbf{Q}\mathbf{K}^{\top}/\sqrt{d}+\mathbf{M}_{\mathrm{causal}})\mathbf{V}\),
\(\sigma\) is the sigmoid, and \(\odot\)
is element-wise multiplication. The value gate \(\mathbf{W}_{gv}\)
attenuates unreliable interaction features before they contribute to
any context, while the output gate \(\mathbf{W}_{go}\) controls how
much the aggregated behavioral context updates each token. Decoupling
token-level reliability filtering from context-level modulation yields
robustness to noisy histories without modifying the attention softmax,
preserving compatibility with efficient attention kernels.

Building on gated-attention designs that primarily modulate the aggregated attention output~\cite{hua2022transformer, qiu2025gated, hstu2024}, DGA additionally gates the value stream, suppressing noisy behavior features before they enter the aggregated context. HSTU uses an unbounded SiLU output gate~\cite{hstu2024}; in our setting, we observe that such gates can exhibit attention-output divergence at scale and therefore adopt bounded sigmoid gates for stable model scaling. Our ablation study shows that the value and output gates are complementary, and Appendix~\ref{app:dga_denoising} further validates the denoising effect of DGA.

\paragraph{Rotary Positional and Temporal Embedding (RoPE + RoTE)}
Language tokens are evenly spaced in an ordinal sequence. Modern LLMs commonly use RoPE~\cite{rope2021}
to encode relative ordinal positions through rotation, a formulation
compatible with FlashAttention-style kernels~\cite{flashattention2022}. User behaviors, however, live in
irregular physical time, where user intent evolves with elapsed time:
two clicks seconds apart and two clicks days apart should be treated
differently even when they occupy adjacent sequence positions. Prior
time-aware recommenders encode time through relative interval embeddings, temporal biases, or bucketized
features~\cite{tisasrec2020, timelyrec2021, hstu2024}, which are less
aligned with length extrapolation and rotary-friendly efficient attention.
We therefore design a rotary temporal embedding that preserves the
rotary formulation while supporting multi-granularity time intervals by
splitting attention heads into two complementary groups.

We propose Rotary Temporal Embedding (RoTE), which rotates queries and keys by the
discretized physical timestamp under a head-specific time granularity \(\tau_h\). To jointly model ordinal order and physical time, we split attention heads into two groups: heads in \(\mathcal{H}_{\mathrm{pos}}\) use standard RoPE~\cite{rope2021} for relative ordinal positions, while heads in \(\mathcal{H}_{\mathrm{time}}\) use RoTE for relative time intervals. Small \(\tau_h\) values at the second/minute scale capture fine temporal proximity, whereas large \(\tau_h\) values at the day/week scale capture coarse recency. For each sequence, we first shift timestamps by its start time, \(\tilde{t}_i=t_i-t_1\), and then define \(b_i=\lfloor \tilde{t}_i/\tau_h \rfloor\). This shift avoids unnecessarily large absolute bucket indices without changing pairwise time intervals. For head \(h\), the attention score between query \(\mathbf{q}_h\) at position \(n\) and key \(\mathbf{k}_h\) at position \(m\) is
\begin{equation}
\label{eq:rope-rote}
\mathrm{Score}_h(\mathbf{q}_h,\mathbf{k}_h)
=
\begin{cases}
\bigl(\mathcal{R}_{\Theta}(n)\mathbf{q}_h\bigr)^{\top}
\bigl(\mathcal{R}_{\Theta}(m)\mathbf{k}_h\bigr),
& h \in \mathcal{H}_{\mathrm{pos}},
\\[4pt]
\bigl(\mathcal{R}_{\Theta}(b_n)\mathbf{q}_h\bigr)^{\top}
\bigl(\mathcal{R}_{\Theta}(b_m)\mathbf{k}_h\bigr),
& h \in \mathcal{H}_{\mathrm{time}},
\end{cases}
\end{equation}
where \(\mathcal{R}_{\Theta}\) is the block-diagonal rotary matrix
and \(b_i\) bucketizes timestamps at granularity \(\tau_h\). Because both position and time are encoded
through the same rotary mechanism, RoTE unifies ordinal and temporal
proximity at multiple granularities within a single attention
operator, which is naturally compatible with FlashAttention.
Appendix~\ref{app:rote_order} derives the relative-time property of RoTE.

\paragraph{Stabilized Residual Normalization (SRN)}
Pure-attention stacks have a theoretical bias toward token uniformity or
rank collapse with depth, although skip connections and MLPs mitigate this
degeneration~\cite{dong2021attention}; related analyses connect over-mixing
and representational collapse to depth in decoder-only LLMs~\cite{barbero2025why}. More generally, the training objective
strongly affects over-smoothing and the utility of deeper Transformer
layers~\cite{xue2023study}. In discriminative
recommendation ranking, however, the sequence encoder is trained from
sparse labels, and the main loss can even be shortcut by
strong non-sequential DLRM features. As a result, directly deepening a
Transformer, even with the LLM-style Pre-LN configuration, often yields
limited gains in recommendation scaling~\cite{guo2024scaling}.

We propose a simple and effective SRN to address this depth-scaling
bottleneck with two minimal modifications. First, it adjusts the LN
placement in a Mix-LN-style depth-dependent manner~\cite{li2025mixln}
to regulate backward gradient flow across layers. Second, it adds a
learnable scalar initialized to a small value to each residual branch to control
how much each layer updates the forward representation~\cite{rezero2021}.
For the \(l\)-th layer (where
\(\mathrm{SubLayer}\in\{\mathrm{Attn},\mathrm{FFN}\}\)),
\begin{equation}
\label{eq:mixln}
\mathbf{x}_{l+1}
=
\begin{cases}
\mathrm{LN}\!\bigl(\mathbf{x}_l + \alpha_l \cdot
\mathrm{SubLayer}(\mathbf{x}_l)\bigr),
& l \le L_{\mathrm{switch}},\\[4pt]
\mathbf{x}_l + \alpha_l \cdot \mathrm{SubLayer}\!\bigl(
\mathrm{LN}(\mathbf{x}_l)\bigr),
& l > L_{\mathrm{switch}},
\end{cases}
\end{equation}
where \(L_{\mathrm{switch}}\) is roughly 25\% of the total depth,
and \(\alpha_l\) is initialized to 0.01. This small initialization makes
the deep encoder start near an identity mapping, so residual updates are
introduced gradually as they receive sufficient training signal. Together
with depth-dependent LN placement, this forward/backward control provides
a more stable residual path under sparse, shortcut-prone supervision and
leads to more consistent gains under depth scaling.

\subsection{Lightweight Cross Decoder (\texorpdfstring{\(C\)}{C})}
\label{subsec:decoder}
The cross decoder $C$ uses a small set of semantic query tokens \(\{\mathtt{[CLS]}, \mathtt{context}, \mathtt{user}, \mathtt{ad}\}\) to attend to the reusable sequence memory from $T$. Since \(C\) runs for every candidate, we keep its activated computation small while preserving candidate-side capacity through two design choices.

\paragraph{Projection-Free Key/Value}
Because \(T\) already produces contextualized memory, \(C\) directly uses \(\mathbf{H}=T(S_u)\) as both keys and values, removing \(\mathbf{W}_K\), \(\mathbf{W}_V\), and the value gate \(\mathbf{W}_{gv}\) from repeated candidate-side computation. Given candidate query tokens \(\mathbf{Z}\), the resulting cross-attention is
\begin{equation}
\label{eq:projection_free_kv}
\mathrm{CrossAttn}(\mathbf{Z},\mathbf{H})
=
\left[
\mathrm{softmax}\!\left(
\frac{(\mathbf{Z}\mathbf{W}_Q)\mathbf{H}^{\top}}{\sqrt{d_h}}
\right)\mathbf{H}
\odot \sigma(\mathbf{Z}\mathbf{W}_{go})
\right]\mathbf{W}_o,
\end{equation}
where \(\mathbf{W}_{go}\) is the output-gate projection.

\paragraph{Token-Specific Parameterization for Query}
The query tokens in \(C\) play different semantic roles, so we assign each token type its own parameters for candidate-side transformations such as \(\mathbf{W}_Q\), \(\mathbf{W}_{go}\) in the cross attention, and FFN sublayers. For a transformation \(\mathcal{F}\), TSP applies
\begin{equation}
\label{eq:tsp}
\mathcal{F}_i(\mathbf{x}_i) = \mathcal{F}(\mathbf{x}_i; \mathbf{W}_i),
\qquad i = 1, \ldots, K.
\end{equation}
Each token activates only its own parameter set, increasing decoder capacity without materially increasing activated FLOPs.

\subsection{Auxiliary Supervision for Sequence Scaling}
\label{subsec:aux-supervision}
Unlike LLMs, industrial recommendation rankers are hybrid systems that fuse non-sequential DLRM features~\cite{dlrm2019} with user behavior sequences~\cite{din2018} in the final prediction head. This creates a training imbalance we call \emph{sequence starvation}: the dense, memorization-heavy non-sequential branch shortcuts the main CVR objective and leaves the sequence branch weakly supervised, hindering effective scaling of the sequence encoder. We address this with two training-only auxiliary objectives: an auxiliary sequence CVR loss that directly supervises the sequence branch, and an alignment loss that regularizes sequence/non-sequence representations.

\paragraph{Auxiliary Sequence CVR Supervision}
Let \(\mathbf{H}_{\mathrm{seq}}\in\mathbb{R}^{K\times d}\) denote the final states of all \(K\) sequence-conditioned query tokens. To provide a direct supervised signal to the sequence side, we attach a lightweight auxiliary sequence CVR head to \(\mathbf{H}_{\mathrm{seq}}\) and the candidate ad features. Unlike the main prediction head, this auxiliary head cannot rely on non-sequential user/context DLRM features. It therefore applies the same BCE supervision to a sequence-conditioned CVR prediction:
\begin{equation}
\label{eq:loss_seq}
\hat{y}_{\mathrm{seq}}
=
\sigma\!\left(g_{\mathrm{seq}}(\mathbf{H}_{\mathrm{seq}}, a)\right),
\qquad
\mathcal{L}_{\mathrm{seq}}
=
\mathbb{E}_{\mathcal{D}}
\big[\ell_{\mathrm{BCE}}(\hat{y}_{\mathrm{seq}}, y)\big].
\end{equation}

\begin{figure}[t]
\centering
\includegraphics[width=0.75\linewidth]{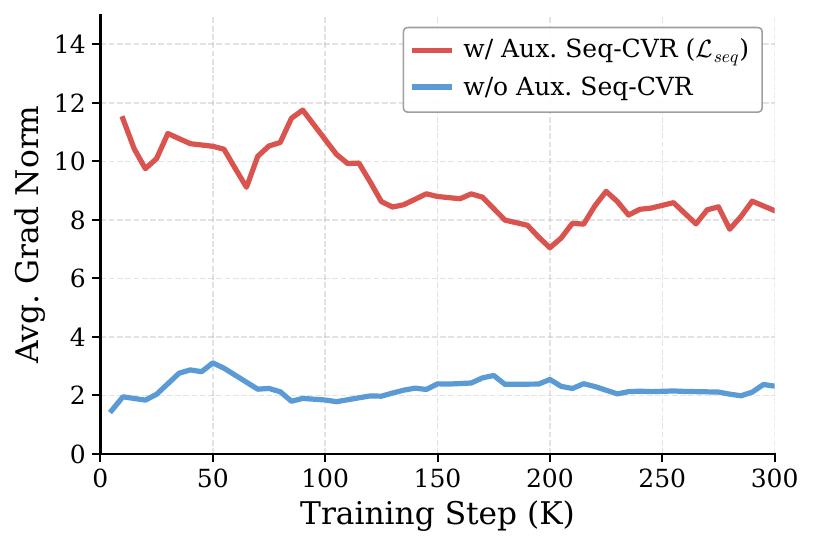}
\caption{Gradient norm of sequence-encoder parameters with and without auxiliary sequence CVR supervision.}
\Description{Line plot comparing average gradient norms of sequence-encoder attention projections with and without auxiliary sequence CVR supervision.}
\label{fig:aux_cvr_gradient}
\end{figure}

As shown in Figure~\ref{fig:aux_cvr_gradient}, even with a small auxiliary weight \(\lambda_{\mathrm{seq}}=0.1\), the auxiliary sequence CVR loss \(\mathcal{L}_{\mathrm{seq}}\) increases the average gradient norm of the sequence encoder by more than \(3\times\), indicating that direct sequence-side supervision substantially strengthens the optimization signal received by the sequence branch. Appendix~\ref{app:gradient_aux_cvr} provides additional details.

\paragraph{Seq/Non-Seq Alignment}
\label{subsec:loss}
Beyond direct label supervision, we regularize the relationship between sequential and non-sequential user representations. Inspired by vision--language representation alignment~\cite{radford2021learning, zhai2023sigmoid}, we treat behavior sequences and non-sequential features as complementary modalities that describe the same user. We therefore align the behavior-sequence and non-sequential representations with a sigmoid contrastive objective, which is less sensitive to batch size than softmax-based contrastive losses and better suited to scaling recommendation models. Concretely, for user \(i\), let \(\mathbf{H}_{\mathrm{seq},i}\) denote the sequence-encoder output and let \(\mathbf{h}_{\mathrm{seq},i}=\mathbf{H}_{\mathrm{seq},i}[\mathtt{CLS}]\) denote its \(\mathtt{[CLS]}\) state, which we use as the sequence-level representation. We feed \(\mathbf{h}_{\mathrm{seq},i}\) and the non-sequential representation into separate two-layer MLP projection heads followed by \(\ell_2\) normalization, obtaining \(\mathbf{z}_{\mathrm{seq},i}\) and \(\mathbf{z}_{\mathrm{ns},i}\), respectively. These projection heads are used only during training and removed at inference. We instantiate the alignment loss as:
\begin{equation}
\label{eq:siglip}
\mathcal{L}_{\mathrm{align}}
=
-\frac{1}{B^2}
\sum_{i=1}^{B}\sum_{j=1}^{B}
\log \sigma\!\left(
y_{ij}\cdot
\bigl(
        \mathbf{z}_{\mathrm{seq},i}^{\top}\mathbf{z}_{\mathrm{ns},j}/\tau - b
\bigr)
\right),
\end{equation}
where \(y_{ij}{=}{+}1\) when the sequence and non-sequence representations come from the same user, and \(-1\) otherwise, with a positive learnable temperature \(\tau\) and a learnable bias \(b\). The overall training objective is:
\begin{equation}
\label{eq:loss_total}
\mathcal{L}_{\mathrm{total}}
=
\mathcal{L}_{\mathrm{ce}}
+
\lambda_{\mathrm{seq}}\mathcal{L}_{\mathrm{seq}}
+
\lambda_{\mathrm{align}}\mathcal{L}_{\mathrm{align}}.
\end{equation}

\subsection{System Co-Design for Computation Reuse}
\label{subsec:system}

To translate \methodname's architectural separation into actual computation reuse, we co-design the system stack around the shared encoder outputs via the following two shared-prefix mechanisms.

\paragraph{User-Level Shared-Prefix Training (ULT)}
In standard instance-level training, samples from the same user often share long overlapping behavior prefixes, causing the sequence encoder $T$ to repeatedly process nearly identical histories. ULT groups samples from the same user within a time window. For a user with $M_u$ samples, we run $T$ once over the union behavior sequence and apply prefix-valid causal masks so that each sample can only attend to its own valid history. This preserves the original chronological training semantics while reducing the encoder computation from $M_u \cdot \mathrm{FLOPs}(T)$ to $\mathrm{FLOPs}(T)$.

\paragraph{Shared-Prefix Serving}
At serving time, a ranking request scores $N$ candidates against the same user history. Inspired by the M-FALCON-style reuse principle~\cite{hstu2024}, we compute the encoder memory $\mathbf{H}=T(S_u)$ once per request and reuse it as shared prefix states. Candidate-side query tokens are then batched and decoded by $C$ against the same memory. The per-request computation changes from
\(N \cdot \mathrm{FLOPs}(T) + N \cdot \mathrm{FLOPs}(C)\) down to
\(\mathrm{FLOPs}(T) + N \cdot \mathrm{FLOPs}(C)\).
Because $C$ is designed to be much lighter than $T$, shared-prefix reuse decouples sequence-side scaling from repeated per-candidate latency, making larger sequence encoders practical under production constraints.

\begin{table*}[!t]
\small
\caption{Comparison with the production DLRM baseline on the large-scale industrial dataset.}
\label{tab:overall}
\begin{center}
\setlength{\tabcolsep}{3.5pt}
\renewcommand{\arraystretch}{1.08}
\begin{tabular}{llccccccc}
\toprule
\textbf{Setting} & \textbf{Metric}
& \textbf{DLRM} & \textbf{DIN} & \textbf{STCA} & \textbf{LLaMA}
& \textbf{HSTU} & \textbf{Trans.} & \textbf{\methodname} \\
\midrule
\multirow{3}{*}{Base}
& AUC $\Delta$ & - & +0.23\% & +0.38\% & +0.53\% & +0.51\% & +0.57\% & \textbf{+0.67\%} \\
& FLOPs (G)    & - & 0.07 & 1.66 & 1.37 & 1.70 & 1.52 & {1.52} \\
& Params (M)   & - & 0.12 & 2.97 & 0.86 & 0.66 & 5.36 & {5.69} \\
\midrule
\multirow{3}{*}{Large}
& AUC $\Delta$ & - & +0.27\% & +0.46\% & +0.70\% & +0.61\% & +0.72\% & \textbf{+0.92\%} \\
& FLOPs (G)    & - & 1.18 & 31.98 & 30.21 & 32.88 & 32.38 & {32.40} \\
& Params (M)   & - & 0.46 & 15.05 & 3.43 & 1.98 & 21.33 & {22.64} \\
\bottomrule
\end{tabular}
\end{center}
\vspace{-1.0em}
\end{table*}

\begin{table*}[!t]
\small
\caption{Public benchmarks on ML-1M, ML-20M, and Amazon-Books.}
\label{tab:public_benchmark}
\begin{center}
\setlength{\tabcolsep}{5pt}
\renewcommand{\arraystretch}{1.08}
\begin{tabular}{llcccccc}
\toprule
\textbf{Dataset} & \textbf{Metric}
& \textbf{DIN} & \textbf{STCA} & \textbf{LLaMA}
& \textbf{HSTU} & \textbf{Trans.} & \textbf{\methodname} \\
\midrule
\multirow{2}{*}{ML-1M}
& AUC      & 0.8064 & 0.8074 & 0.8087 & 0.8079 & 0.8090 & \textbf{0.8099} \\
& Log loss & 0.5346 & 0.5372 & 0.5334 & 0.5343 & 0.5329 & \textbf{0.5317} \\
\midrule
\multirow{2}{*}{ML-20M}
& AUC      & 0.8070 & 0.8079 & 0.8088 & 0.8082 & 0.8103 & \textbf{0.8135} \\
& Log loss & 0.5329 & 0.5333 & 0.5317 & 0.5322 & 0.5302 & \textbf{0.5292} \\
\midrule
\multirow{2}{*}{Books}
& AUC      & 0.7521 & 0.7527 & 0.7541 & 0.7566 & 0.7556 & \textbf{0.7571} \\
& Log loss & 0.4345 & 0.4357 & 0.4305 & 0.4242 & 0.4244 & \textbf{0.4221} \\
\midrule
& {FLOPs (M)}  & 1.90 & 24.10 & 21.13 & 29.77 & 23.32 & {23.41} \\
& {Params (M)} & 0.01 & 0.11 & 0.03 & 0.02 & 0.17 & {0.18} \\
\bottomrule
\end{tabular}
\end{center}
\vspace{-0.8em}
\end{table*}

\section{Experiments}
\label{sec:exp}

We conduct comprehensive experiments to evaluate \methodname's effectiveness, scalability, efficiency, and deployability. Specifically, we answer five research questions:
(RQ1): Does \methodname outperform other sequence models under production FLOPs budgets?
(RQ2): Does \methodname scale better in length, depth, and width than LLM-style baselines?
(RQ3): How much does each \methodname design component contribute?
(RQ4): Do user-level shared-prefix training (ULT) and shared-prefix serving improve training and serving efficiency?
(RQ5): Does \methodname improve online business metrics under the P99 latency budget?

\subsection{Experimental Setup}
We primarily evaluate \methodname on a large-scale industrial dataset from TikTok Shop Ads. All data are strictly anonymized and de-identified before use, containing no personally identifiable information. We keep the ranking stack, behavior sequence, and query tokens fixed across models, and replace only the sequence-modeling component. Due to confidentiality constraints, we report the relative AUC improvement (AUC $\Delta$) over a highly optimized production DLRM baseline for the industrial evaluation. We also evaluate on three public benchmarks: MovieLens-1M and MovieLens-20M~\cite{harper2015movielens}, and Amazon-Books~\cite{mcauley2015image}. To align these public benchmarks with our primary industrial CVR task, we cast their standard sequential recommendation protocols into a pointwise prediction task via chronological splitting. For these public datasets, we report absolute AUC and log loss under identical evaluation protocols for all models, with additional dataset and implementation details provided in the Appendix.

We group baselines into three families. (1) Target-aware attention baselines include DIN~\cite{din2018} and Stacked Target-to-History Cross Attention (STCA)~\cite{guan2026make}. (2) Causal sequence-model baselines include a LLaMA-style Transformer~\cite{llama2023} and HSTU~\cite{hstu2024}. (3) Trans., a strong encoder--decoder baseline, uses a modernized bidirectional encoder following ModernBERT~\cite{warner2025smarter}, with Pre-LN, GeGLU, and RoPE\@. We equip Trans.\ with token-specific parameterization (TSP) and configure it to approximately match the parameter count of \methodname{}. For a fair comparison, all models use the same sequence preprocessing pipeline and receive the same preprocessed behavior sequence as input. Reported FLOPs and parameter counts cover only the sequence-modeling component. Unless otherwise specified, all sequence-modeling baselines use the auxiliary sequence CVR loss during training to mitigate sequence starvation.

\subsection{Overall Comparison (RQ1)}

\begin{figure}[h]
    \centering
    \includegraphics[height=0.24\textheight,width=\linewidth,keepaspectratio]{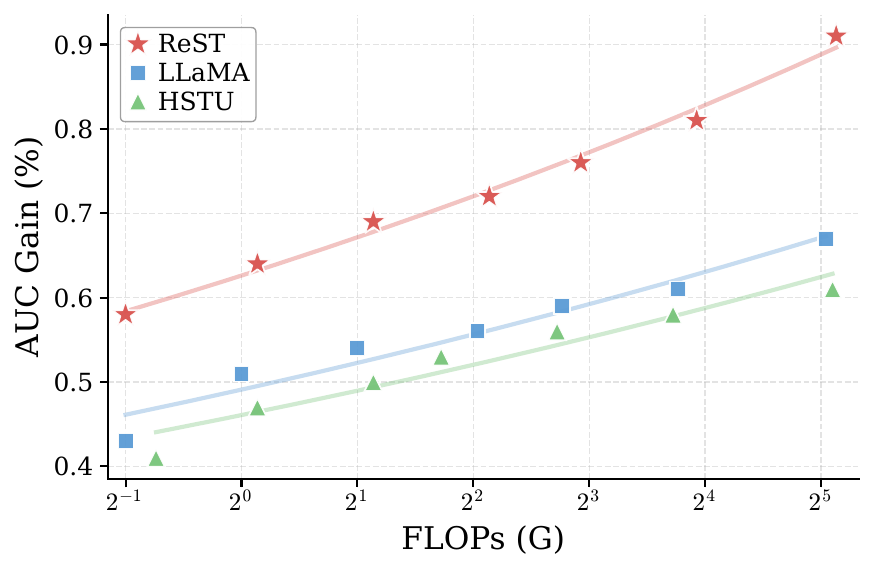}
    \caption{Accuracy vs FLOPs on the internal dataset. \methodname achieves favorable accuracy-FLOPs scaling.}
    \Description{Plot of AUC improvement versus FLOPs comparing \methodname with sequence-modeling baselines on the industrial dataset.}
    \label{fig:joint_scale}
\end{figure}

\begin{figure*}[!t]
    \centering
    \begin{minipage}[t]{0.32\textwidth}
        \centering
        \includegraphics[width=\linewidth]{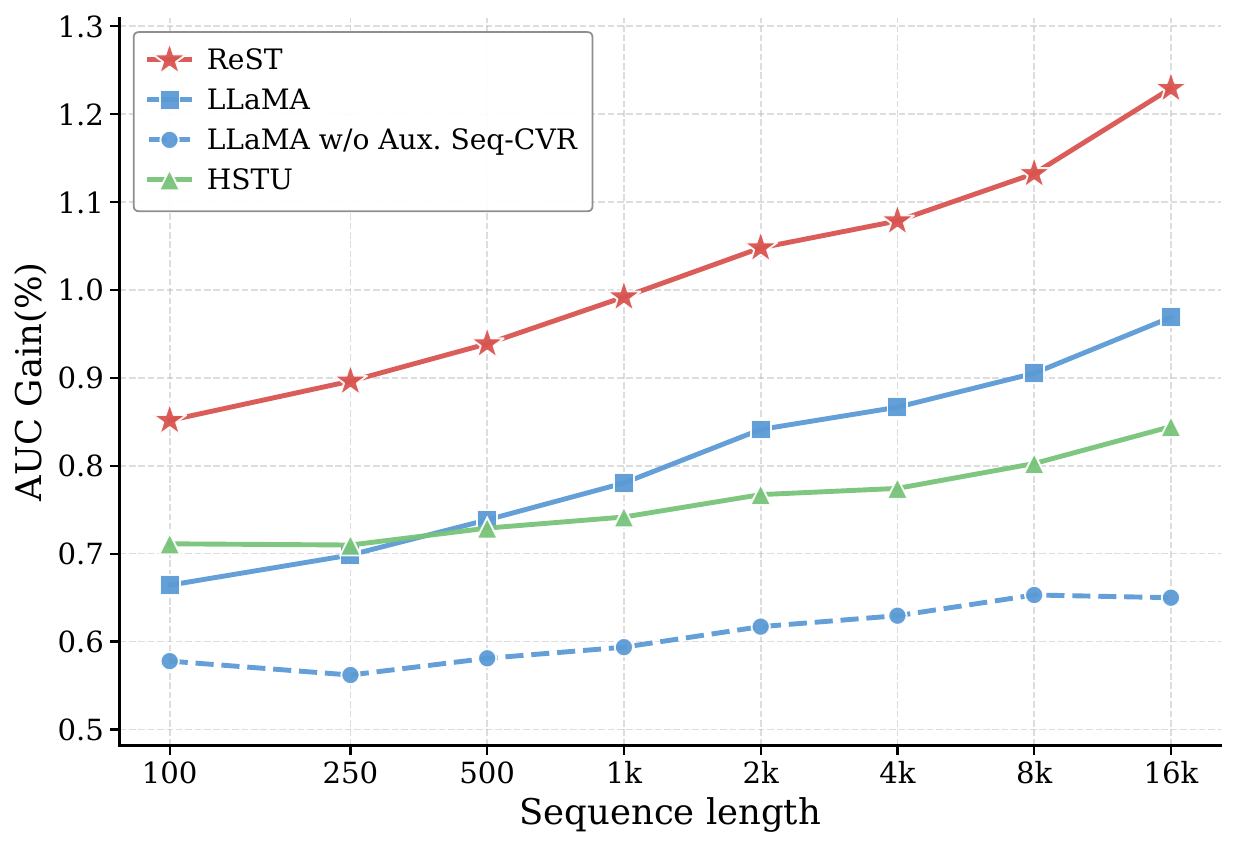}
        \vspace{0.2em}

        {\footnotesize (a) Sequence Scale}
    \end{minipage}
    \hfill
    \begin{minipage}[t]{0.32\textwidth}
        \centering
        \includegraphics[width=\linewidth]{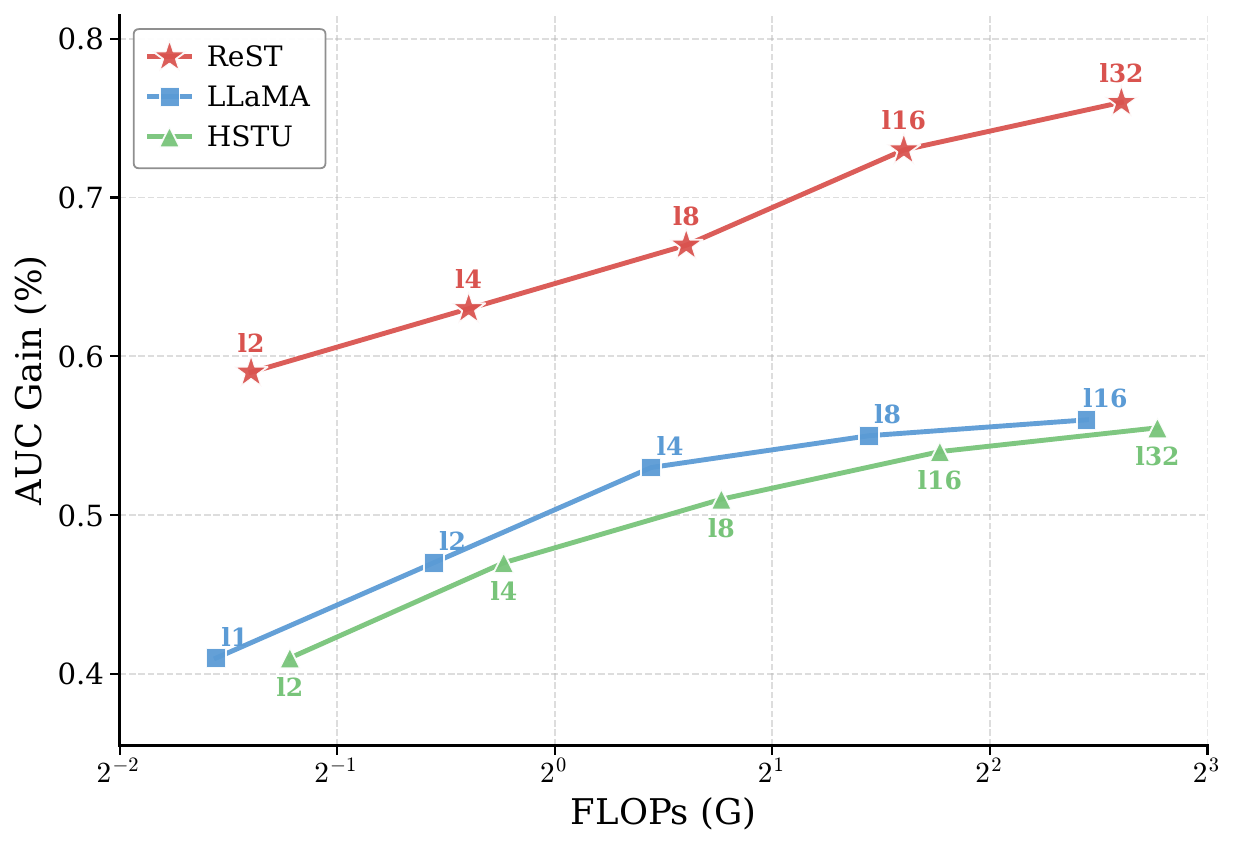}
        \vspace{0.2em}

        {\footnotesize (b) Depth Scale}
    \end{minipage}
    \hfill
    \begin{minipage}[t]{0.32\textwidth}
        \centering
        \includegraphics[width=\linewidth]{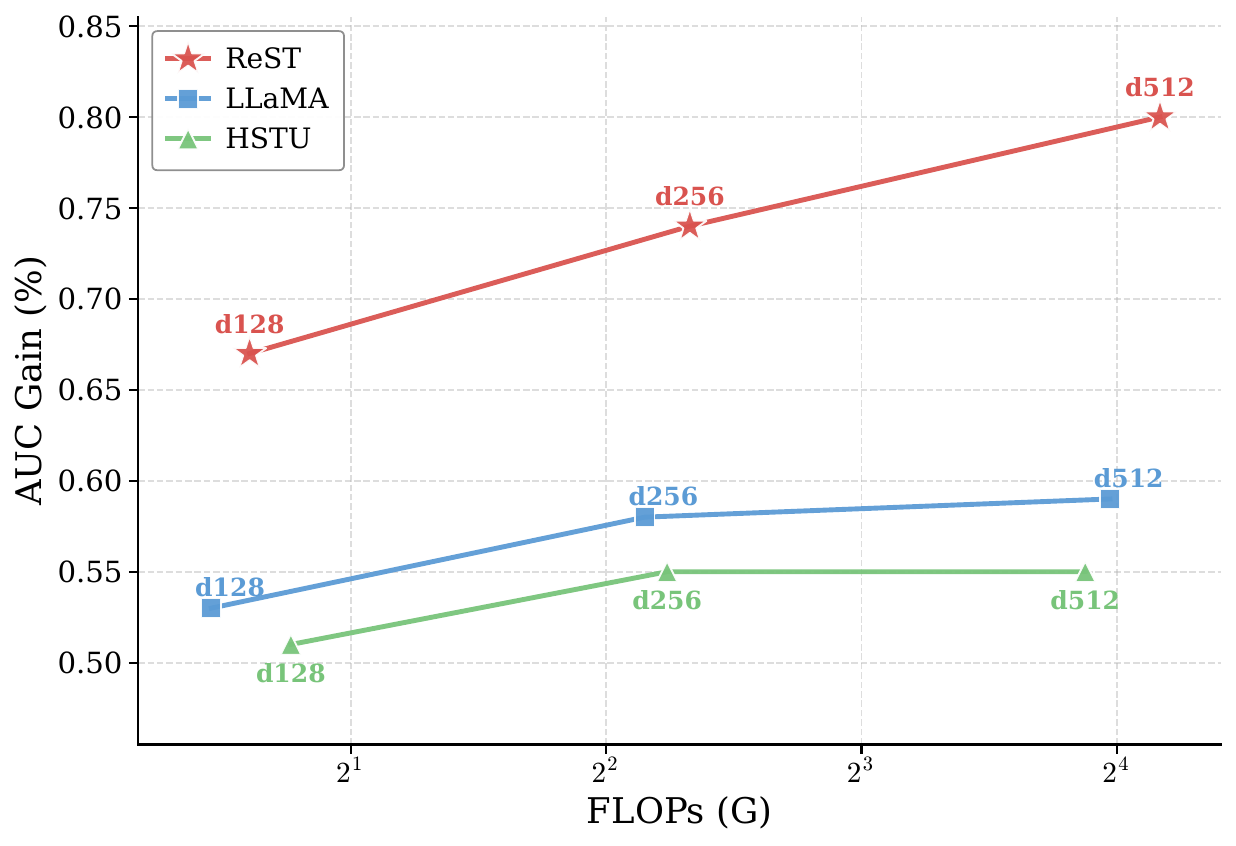}
        \vspace{0.2em}

        {\footnotesize (c) Width Scale}
    \end{minipage}
    \caption{Scaling behavior along three axes: (a) sequence length, (b) model depth, and (c) hidden dimension. \methodname consistently outperforms competing sequence models across the explored budgets.}
    \Description{Three subplots showing scaling trends for sequence length, model depth, and hidden dimension across \methodname and baseline sequence models.}
    \label{fig:scaling}
\end{figure*}

Table~\ref{tab:overall} reports the main comparison on the large-scale industrial dataset. We replace the sequence-modeling component in the production DLRM ranker and roughly align FLOPs across the main sequence-scaling models. \methodname consistently outperforms all baseline families at comparable FLOPs, reaching the best AUC lift under both Base and Large budgets and \(+0.92\%\) in the Large setting. In this production ranking setting, an offline AUC \(\Delta\) of \(0.05\%\) is considered practically meaningful and can translate into measurable online gains. Among the baselines, LLaMA and HSTU outperform DIN and STCA, confirming the value of richer sequence-side self-attention. HSTU is weaker than LLaMA, suggesting that scaling recipes for generative recommendation may not transfer directly to discriminative ranking. The modern encoder--decoder Trans.\ further improves over LLaMA and is approximately parameter-matched to the full \methodname, yet still trails it by \(0.20\%\) AUC in the Large setting. Table~\ref{tab:abl_struct} provides a second parameter-controlled comparison: the parameter-rich cross decoder \(C\) contributes \(+0.05\%\) AUC, while the encoder-only \methodname, whose parameter count is close to that of LLaMA, retains approximately \(+0.87\%\) AUC lift and outperforms LLaMA by \(0.17\%\). Together, these comparisons show that \methodname's gains arise from its recommendation-native sequence modeling and asymmetric architecture rather than parameter count alone.

We then report results on three public datasets in Table~\ref{tab:public_benchmark}. Consistent with the industrial findings, \methodname achieves the best AUC and log loss across all datasets, while maintaining comparable FLOPs. This suggests that the benefits of \methodname are not merely artifacts of large-scale industrial data, but also transfer to standard public benchmarks.

\subsection{Scaling Analysis (RQ2)}

\noindent\textbf{Overall scaling.} Figure~\ref{fig:joint_scale} visualizes the joint scaling behavior when layer depth, hidden dimension, and sequence length are scaled together. We focus this scaling analysis on scalable self-attention-style sequence models, since their sequence-side computation can be cached, making them the most relevant family for long-context scaling under production ranking constraints. As a descriptive summary over the tested compute range, we fit $\Delta\text{AUC}(F) = a \cdot F^{b}$, where \(F\) denotes FLOPs, and obtain $(a, b) = (0.626, 0.101)$ for \methodname, $(0.491, 0.090)$ for LLaMA, and $(0.461, 0.088)$ for HSTU, with log-space $R^2 = 0.991$, $0.935$, and $0.924$, respectively. \methodname attains consistently higher AUC gains across the tested compute regime, suggesting that rec-native designs yield a more favorable empirical compute-performance trend in our setting. This advantage becomes clearer in the axis-wise scaling analysis below: when scaling context length, depth, or width independently, LLM-style sequence models show earlier diminishing returns, whereas \methodname continues to convert additional sequence-side computation into accuracy gains.

\noindent\textbf{Sequence length scale.} Figure~\ref{fig:scaling}(a) varies the available behavior context from 100 to 16k tokens while keeping depth and hidden dimension fixed. We evaluate each context-length setting on recent data with sufficient history coverage and report each model's AUC lift over the production baseline evaluated on the same samples. Without auxiliary sequence CVR supervision, LLaMA exhibits the sequence-starvation effect discussed in Sec.~\ref{subsec:aux-supervision}: extending the available context yields almost no additional lift. With auxiliary sequence CVR supervision, all sequence models show stronger gains at larger context budgets, with the ordering \methodname \(>\) LLaMA \(>\) HSTU\@. Across the evaluated settings, \methodname continues to improve up to 16k tokens, achieving an additional AUC lift of approximately \(0.4\%\) over the evaluated context range. This represents the largest observed increase among the compared models and suggests that rec-native designs enable more effective use of longer user histories.

\noindent\textbf{Model depth scale.} Figure~\ref{fig:scaling}(b) varies the number of layers while keeping sequence length and hidden dimension fixed. For \methodname, $T$ and $C$ depths are scaled 1:1. As depth increases, \methodname improves steadily, with AUC gain rising from $+0.59\%$ to $+0.76\%$. In contrast, LLaMA and HSTU improve at shallow depths but exhibit diminishing returns beyond 8 and 16 layers, respectively. This pattern suggests that simply stacking deeper generic sequence blocks is insufficient for recommendation ranking: under sparse supervision and noisy behavior data, additional depth may saturate before being effectively optimized. \methodname benefits more consistently from depth, and the component ablation in Sec.~\ref{sec:ablation} further confirms that SRN is especially important for deeper models.

\noindent\textbf{Hidden dimension scale.} Figure~\ref{fig:scaling}(c) scales the hidden dimension $d \in \{128, 256, 512\}$ with depth and sequence length fixed. Both baselines saturate early: HSTU and LLaMA flatten around $d{=}256$, and increasing to $d{=}512$ yields almost no additional AUC despite a substantial increase in FLOPs. \methodname follows a different trend, continuing to improve from $+0.67\%$ to $+0.80\%$ and remaining the only tested model with monotonic gains at the largest width. The combination of gated attention, temporal encoding, and stabilized residual learning allows \methodname to benefit from larger hidden dimensions rather than merely increasing arithmetic cost.

\begin{table}[!t]
\centering
\small
\caption{Ablation: architecture and rec-native block designs.}
\label{tab:abl_struct}
\setlength{\tabcolsep}{5pt}
\begin{tabular}{@{}llccc@{}}
\toprule
\textbf{Group} & \textbf{Variant} & \textbf{AUC $\Delta$} & \textbf{Param $\Delta$} & \textbf{FLOPs $\Delta$} \\
\midrule
\multirow{5}{*}{\textbf{Architecture}}
& T   & base    & ---      & ---  \\
  & + MoE in $T$                                  & +0.00\% & +500\% & +0.2\%  \\
& + C (w/o TSP)                         & +0.03\% & +100.3\% & +0.2\%  \\
  & + C (w/o PF-KV)                         & +0.05\% & +473.2\% & +6.7\%  \\
  & \textbf{+ C}           & \textbf{+0.05\%} & \textbf{+473.2\%} & \textbf{+0.2\%} \\

\midrule
\multirow{4}{*}{\textbf{Gated Attn}}
    & w/o gate     & base    & ---     & ---     \\
  & + O gate                               & +0.03\% & +7.6\%  & +3.7\%  \\
  & + O \& QKV gate                        & +0.06\% & +11.4\% & +14.9\% \\
  & \textbf{+ O \& V gate}                 & \textbf{+0.06\%} & \textbf{+8.9\%} & \textbf{+7.5\%} \\
\midrule
\multirow{4}{*}{\textbf{Norm}}
  & Pre-LN & base & --- & --- \\
  & Post-LN & +0.00\% & \(\approx 0\) & \(\approx 0\) \\
  & Mix-LN & +0.02\% & \(\approx 0\) & \(\approx 0\) \\
  & \textbf{SRN} & \textbf{+0.08\%} & \(\approx 0\) & \(\approx 0\) \\
\midrule
\multirow{4}{*}{\textbf{Pos./Temporal}}
  & w/o pos & base & --- & --- \\
  & RoPE & +0.04\% & \(\approx 0\) & \(\approx 0\) \\
& RoPE + RoTE (sec) & +0.08\% & \(\approx 0\) & \(\approx 0\) \\
  & \textbf{RoPE + RoTE} & \textbf{+0.12\%} & \(\approx 0\) & \(\approx 0\) \\
\bottomrule
\end{tabular}
\end{table}

\subsection{Ablation Studies (RQ3)}
\label{sec:ablation}
\noindent\textbf{Architecture and Compute Allocation.}
We study where additional capacity should be allocated in \methodname.
As shown in the Architecture group of Table~\ref{tab:abl_struct}, simply adding MoE-based capacity~\cite{wang2024auxiliary} to the sequence encoder $T$ increases parameters by $500\%$ but yields no measurable AUC gain, suggesting that encoder capacity alone is not the limiting factor. In contrast, adding a lightweight cross decoder \(C\) improves AUC by $+0.03\%$ with only $+0.2\%$ extra FLOPs, showing the value of candidate-conditioned readout. Token-specific parameterization further raises the gain to $+0.05\%$, while projection-free KV keeps the FLOPs overhead at $+0.2\%$ rather than $+6.7\%$. Thus, \methodname benefits from allocating capacity to the heterogeneous candidate-conditioned decoder while keeping repeated per-candidate arithmetic nearly unchanged.

\noindent\textbf{Gated Attention.}
The Gated Attention group of Table~\ref{tab:abl_struct} shows that an output gate alone already yields a $+0.03\%$ gain. Adding gates to all Q/K/V branches increases AUC to $+0.06\%$, but also incurs a large FLOPs overhead ($+14.9\%$). In contrast, combining the output gate with a value gate reaches the same $+0.06\%$ gain at roughly half the extra compute ($+7.5\%$ FLOPs). Appendix~\ref{app:dga_denoising} further shows that combining the output and value gates yields better performance under noisy behavior perturbations. We therefore adopt Output + Value gating as our Dual-Gated Attention design for a better accuracy--efficiency trade-off.

\begin{table}[t]
\centering
\small
\caption{Ablation: auxiliary objectives.}
\label{tab:aux_supervision}
\setlength{\tabcolsep}{7pt}
\begin{tabular}{lccc}
\toprule
\textbf{Backbone} & \(\mathcal{L}_{\mathrm{seq}}\) {(Base)} & \(\mathcal{L}_{\mathrm{seq}}\) {(Large)} & \(\mathcal{L}_{\mathrm{align}}\) \\
\midrule
DIN & +0.02\% & +0.02\% & --- \\
STCA & +0.13\% & +0.20\% & --- \\
LLaMA & +0.16\% & +0.22\% & +0.05\% \\
HSTU & +0.10\% & +0.15\% & +0.05\% \\
\methodname & +0.07\% & +0.12\% & +0.06\% \\
\bottomrule
\end{tabular}
\end{table}

\noindent\textbf{Normalization and Residual Connections.}
We then evaluate how normalization placement and residual scaling affect optimization in a fixed 16-layer setting. As shown in the Norm group of Table~\ref{tab:abl_struct}, the LLaMA-style Pre-LN design serves as the local baseline; switching to Post-LN brings no gain, while Mix-LN provides a modest $+0.02\%$ improvement, suggesting that normalization placement matters for deeper recommendation models. Adding scaled residual learning on top of this design, SRN achieves the largest gain of $+0.08\%$, substantially outperforming the other variants at the same depth. This indicates that training deeper sequence models on sparse recommendation data benefits from both appropriate normalization placement and controlled residual updates.

\noindent\textbf{Positional and Temporal Encoding.}
The Positional and Temporal Encoding group in Table~\ref{tab:abl_struct} starts from a no-position baseline. Adding RoPE improves AUC by $+0.04\%$, confirming the benefit of ordinal positional information. Adding single-granularity RoTE with one-second buckets on top of RoPE raises the gain to $+0.08\%$, showing that physical time provides complementary signal beyond ordinal position. The default multi-granularity RoTE, which assigns distinct temporal granularities to different temporal attention heads, further raises the gain to $+0.12\%$, demonstrating the benefit of modeling temporal proximity at multiple resolutions. Importantly, all variants already receive the same delta-time features as token-level side information, so these gains reflect the additional value of injecting physical time directly into the attention mechanism rather than merely exposing the model to temporal features. Further details of RoTE are provided in Appendix~\ref{app:rote_order}.

\noindent\textbf{Auxiliary Objectives.}
Table~\ref{tab:aux_supervision} summarizes the effects of the two training-only auxiliary objectives. Auxiliary sequence CVR supervision improves all sequence-aware backbones, with larger gains in the Large setting: STCA improves from \(+0.13\%\) to \(+0.20\%\), LLaMA from \(+0.16\%\) to \(+0.22\%\), HSTU from \(+0.10\%\) to \(+0.15\%\), and \methodname from \(+0.07\%\) to \(+0.12\%\). This supports our motivation that larger sequence modules require stronger direct sequence-side supervision to avoid being shortcut by the dense non-sequential branch. The smaller marginal gain on \methodname suggests that its rec-native encoder is less dependent on the auxiliary objective, although auxiliary supervision remains beneficial. On top of auxiliary sequence CVR supervision, the alignment objective further improves LLaMA, HSTU, and \methodname by $+0.05\%$, $+0.05\%$, and $+0.06\%$, respectively, indicating that sequence/non-sequence alignment provides a stable additional regularization signal across scalable sequence models. Since these objectives are architecture-agnostic and introduce no serving-time overhead, we enable auxiliary sequence CVR supervision for all sequence-scaling baselines in the main comparison.

\subsection{System Efficiency and Online A/B Test Result (RQ4 \& RQ5)}
To translate the reuse boundary of \methodname into actual system efficiency, we implement the training and serving stack around the shared sequence encoder. On the training side, our baseline is already a highly optimized pipeline with fused attention/GEMM kernels, mixed-precision training, and padding removal for variable-length sequences. On top of this strong baseline, user-level shared-prefix training (ULT) further improves end-to-end training throughput by $5.8\times$ by removing redundant prefix computation across samples from the same user, showing that the gain comes from computation reuse rather than from low-level kernel optimization alone. On the serving side, request-level shared-prefix serving is the key enabler for online sequence scaling: $T$ is executed once per request and reused across all candidates, yielding up to a $20\times$ reduction in the per-request inference cost of the sequence-modeling component. This reuse is critical for deploying \methodname online under strict production latency budgets and substantially narrows the deployment gap between large sequence models and industrial ranking systems.

\begin{table}[h]
\centering
\small
\caption{Online A/B results.}
\label{tab:online_ab}
\setlength{\tabcolsep}{6pt}
\begin{tabular}{@{}ccc@{}}
\toprule
\textbf{AUC $\Delta$} & \textbf{Advv Lift} & \textbf{\(p\)-value}\\
\midrule
+1.31\% & +11.93\% & $<0.01$ \\
\bottomrule
\end{tabular}
\end{table}

To validate the practical impact of \methodname, we conduct a one-week online A/B test on TikTok Shop Ads, one of the largest advertising platforms. Users are randomly assigned to control and treatment groups, with the treatment group receiving 20\% of production traffic. The control is a strong production DLRM ranker, while the treatment replaces its sequence modeling component with \methodname. We scale \methodname to a deployable configuration that satisfies the $50$\,ms P99 latency budget. We use a revenue-related business metric, Advertiser Value (Advv), as the primary online metric, report streaming AUC for online prediction quality, and monitor end-to-end P99 latency and other production guardrails. As shown in Table~\ref{tab:online_ab}, this latency-constrained configuration improves AUC by $+1.31\%$ and lifts Advv by a statistically significant $+11.93\%$ ($p<0.01$), while satisfying all other monitored production guardrails. \methodname has been fully deployed in production, showing that scaling behavior-sequence modeling through rec-native designs can translate offline gains into substantial business value in a realistic large-scale production setting.

\section{Related Work}
\label{sec:related}

\noindent\textbf{Deep Learning Recommendation Models (DLRMs).}
Modern industrial ranking systems are largely built on the DLRM
paradigm~\cite{cheng2016wide, dlrm2019}, which combines large embedding tables
with relatively lightweight interaction modules. A long line of
work has refined this paradigm from complementary angles.
DCN-V2~\cite{dcnv2} strengthens explicit feature crossing; DIN~\cite{din2018} introduces target-aware attention for user-behavior modeling; and SIM/SDIM~\cite{sim2020,sdim2022} extend behavior modeling to much longer histories through retrieval-based approximation. Together, these methods establish behavior sequences as highly informative signals for ranking, but typically keep sequence modeling compact or heavily compressed within a larger ranking stack rather than treating it as a primary scaling axis.

\noindent\textbf{Large Recommendation Models (LRMs).}
Recent work on large recommendation models can be viewed through
the lens of what is being scaled. One line scales the joint
interaction between behavior sequences and dense non-sequential
features: RankMixer~\cite{rankmixer2024} scales dense feature
interaction through hardware-aware token mixing; OneTrans~\cite{onetrans2023}
and InterFormer~\cite{interformer2023} integrate behavior sequences with
non-sequential features; and FAT~\cite{fat2023} and
MTmixAtt~\cite{mtmixatt2023} scale heterogeneous feature interaction
through field-aware or mixture-of-experts parameterization. These methods strengthen the
joint feature-interaction stack rather than isolating the behavior
sequence as the main scaling object, making them complementary to
our focus. A second line scales sequence modeling more directly. Building on sequence Transformers such as SASRec~\cite{sasrec2018}, STCA~\cite{guan2026make} strengthens target-to-history cross attention, LLaMA-style decoder-only baselines~\cite{llama2023, guo2024scaling} enlarge sequence-model capacity with generic self-attention blocks, and HSTU~\cite{hstu2024} demonstrates strong scaling in a generative recommendation formulation. Yet these recipes do not fully address behavior-sequence scaling in discriminative ranking, where noisy and sparsely supervised histories must be scored against many candidates under strict latency constraints. \methodname{} addresses these limitations by making the behavior sequence a reusable primary scaling axis through rec-native designs.

\section{Conclusion}
\label{sec:conclusion}
In this paper, we introduced \methodname, a recommendation-native framework for scaling behavior-sequence Transformers in industrial ranking under strict latency constraints. By addressing the signal quality and computation asymmetry challenges through rec-native sequence modeling, asymmetric architecture design, and shared-prefix computation reuse, \methodname mitigates sequence starvation and scales more effectively along behavior sequence length, depth, and width than LLM-style blocks. This enables larger sequence models to achieve substantial business gains under practical latency budgets, leading to full production deployment. Our results suggest that, with rec-native designs, behavior-sequence scaling remains an under-exploited axis for industrial ranking and can translate into substantial online business value.

\begin{acks}
We thank Chenzhi Zhou for his valuable PMO support, especially in facilitating cross-functional collaboration and production deployment.
\end{acks}

\bibliographystyle{ACM-Reference-Format}
\bibliography{references}

\appendix

\section{Appendix}
\label{app:public_details}

\subsection{Denoising Analysis of Dual-Gated Attention}
\label{app:dga_denoising}
To further validate the denoising role of Dual-Gated Attention (DGA), we conduct a controlled noise-injection study. Specifically, we inject Gaussian noise into behavior tokens at different noise levels and compare standard attention, output-gated attention, and DGA under the same training and evaluation protocol. This setup isolates whether the attention block can remain robust when a growing fraction of behavior evidence becomes unreliable.

\begin{figure}[t]
\centering
\includegraphics[width=0.9\linewidth]{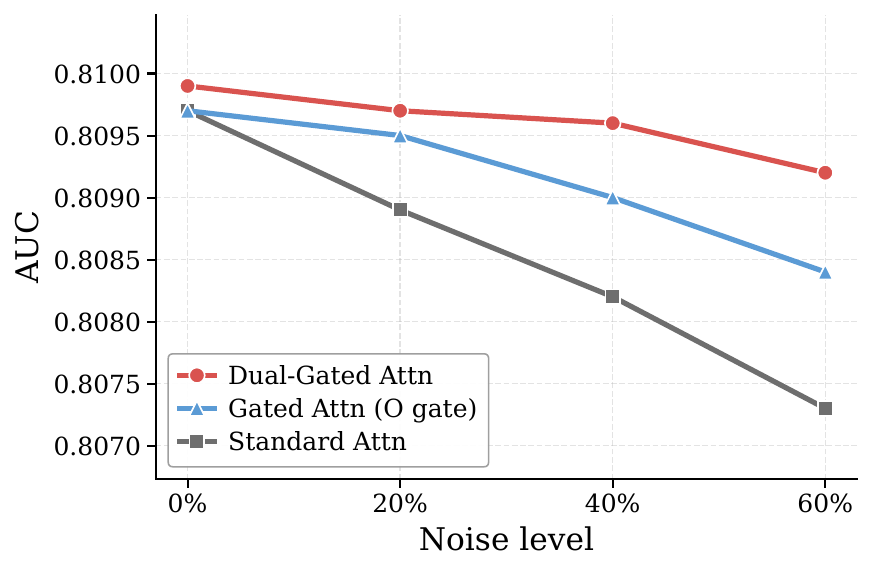}
\caption{Dual-Gated Attention is more robust as token noise increases, supporting its value-side filtering and output-side modulation design.}
\Description{Plot comparing standard attention, output-gated attention, and Dual-Gated Attention under different synthetic token-noise levels on MovieLens-1M.}
\label{fig:gate_attn_denoising}
\end{figure}

Figure~\ref{fig:gate_attn_denoising} shows that DGA is consistently more robust as token noise increases. Standard attention degrades most rapidly because noisy behavior features directly enter the value stream and are aggregated into the contextual representation. Output-gated attention is more stable, but still cannot prevent corrupted token features from being mixed during aggregation. By contrast, DGA gates both the value stream before aggregation and the output stream after aggregation, allowing the model to suppress unreliable evidence before it contaminates the sequence context. This result supports our design choice that value and output gates are complementary for noisy recommendation sequences.

\subsection{Relative-Time Property of RoTE}
\label{app:rote_order}
RoTE follows the same principle as RoPE: query and key vectors are rotated according to their timestamp buckets, but their dot product cancels the shared coordinate frame and depends only on the relative rotation. We provide a compact derivation for a temporal head. As defined in the main text, timestamps are shifted by the sequence start time before bucketization: \(\tilde{t}_i=t_i-t_1\) and \(b_i=\lfloor \tilde{t}_i/\tau_h \rfloor\). This keeps bucket indices within the time span of each sequence rather than the range of absolute wall-clock timestamps. Let \(b_n\) and \(b_m\) denote the resulting buckets under head-specific granularity \(\tau_h\). The RoTE attention score can be written as
\begin{align}
\mathrm{Score}_{\mathrm{RoTE}}(\mathbf{q}, \mathbf{k})
&=
\left(\mathcal{R}_{\Theta}(b_n)\mathbf{q}\right)^{\top}
\left(\mathcal{R}_{\Theta}(b_m)\mathbf{k}\right) \nonumber\\
&=
\mathbf{q}^{\top}
\mathcal{R}_{\Theta}(b_n)^{\top}
\mathcal{R}_{\Theta}(b_m)
\mathbf{k} \nonumber\\
&=
\mathbf{q}^{\top}
\mathcal{R}_{\Theta}(-b_n)
\mathcal{R}_{\Theta}(b_m)
\mathbf{k} \nonumber\\
&=
\mathbf{q}^{\top}
\mathcal{R}_{\Theta}(b_m-b_n)
\mathbf{k}.
\label{eq:rote_relative_appendix}
\end{align}
The derivation uses two basic properties of rotation matrices:
\[
\mathcal{R}_{\Theta}(x)^{\top}=\mathcal{R}_{\Theta}(-x),
\qquad
\mathcal{R}_{\Theta}(x)\mathcal{R}_{\Theta}(y)=\mathcal{R}_{\Theta}(x+y).
\]
Therefore, the temporal head does not depend on the shifted timestamp buckets \(b_n\) and \(b_m\) separately, but only on their relative bucket difference \(b_m-b_n\). Because subtracting the sequence start time cancels in pairwise differences, it does not change the elapsed time represented between two tokens.

This derivation establishes relative-time dependence, not a monotonic or one-to-one mapping from elapsed time to attention score: rotary features are periodic, and the resulting score also depends on the query and key contents. Events within the same bucket are intentionally indistinguishable at that granularity. Assigning different \(\tau_h\) values to different heads exposes relative time intervals at complementary resolutions while retaining compatibility with the rotary computation used by efficient attention kernels. In our industrial implementation, we reserve one attention head for RoPE and assign the remaining heads to head-specific temporal granularities $\tau_h \in \{\mathrm{second},\, \mathrm{minute},\, \mathrm{hour},\, \mathrm{day},\, \mathrm{week},\, \mathrm{quarter},\, \mathrm{year}\}$, enabling different temporal heads to capture recency at complementary resolutions, from fine-grained second/minute patterns to long-horizon quarter/year trends.

\subsection{Gradient Analysis for Auxiliary Sequence CVR Supervision}
\label{app:gradient_aux_cvr}
To further diagnose the effect of auxiliary sequence CVR supervision, we monitor the gradient norm received by the Transformer sequence encoder during training. As discussed in Section~\ref{subsec:aux-supervision}, the main CVR objective is optimized after sequence representations are fused with strong non-sequential DLRM features, so the prediction head can partially reduce the loss through the non-sequential branch. This shortcut can leave the sequence encoder with weak gradients, especially when the sequence module is scaled up.

Figure~\ref{fig:aux_cvr_gradient} reports the average gradient norm of projection matrices across all attention layers in the sequence encoder. Specifically, at each training checkpoint, we compute the gradient norms of the $Q$, $K$, and $V$ projection matrices in every sequence-encoder attention layer and report their average across layers. Even with a small auxiliary weight \(\lambda_{\mathrm{seq}}=0.1\), auxiliary sequence CVR supervision increases the average gradient norm by more than \(3\times\) (from \(2.35\) to \(8.50\)), indicating that direct sequence-side supervision substantially strengthens the optimization signal received by the sequence branch. We observe a similar trend for the FFN parameters, suggesting that the auxiliary objective improves gradient flow throughout the sequence encoder rather than only in the attention projections.

\subsection{Implementation Details}
\label{app:public_implementation}
\noindent\textbf{Public Benchmark Settings.}
All public benchmarks are used only for offline academic benchmarking; models are implemented with RecBole~\cite{zhao2021recbole} and trained using Adam. We tune the learning rate and weight decay over \(\{10^{-4}, 5{\times}10^{-4}, 10^{-3}\}\), use a batch size of 4,096, and train for up to 20 epochs with early stopping on validation AUC\@. We apply 5-core filtering to each dataset, retaining only users and items with at least five interactions. Table~\ref{tab:dataset_overview} summarizes the resulting dataset statistics.
\begin{table}[h]
\centering
\small
\caption{Public dataset overview.}
\label{tab:dataset_overview}
\setlength{\tabcolsep}{4.5pt}
\begin{tabular}{lccc}
\toprule
\textbf{Statistic} & \textbf{ML-1M} & \textbf{ML-20M} & \textbf{Books} \\
\midrule
Users & 6K & 138K & 604K \\
Items & 4K & 18K & 368K \\
Interactions & 1M & 20M & 9M \\
\bottomrule
\end{tabular}
\end{table}

\noindent\textbf{Evaluation of Public Benchmarks.}
Ratings of 4 or above are treated as positive samples and lower ratings as negative samples. No additional negative sampling is performed; the observed low-rating interactions serve as negative instances. For each user, interactions are sorted chronologically and split into train/validation/test sets with an 80\%/10\%/10\% ratio. We report AUC and binary log loss under this labeled binary prediction setting.

\noindent\textbf{Model Configuration of Public Benchmarks.}
The maximum behavior sequence length is 200 and the item embedding dimension is 32. Transformer-based models use two sequence-modeling layers and, for encoder--decoder architectures, two additional decoder layers, with two attention heads and a feed-forward inner dimension of 128. The prediction head is a three-layer MLP with hidden sizes \([256,128,64]\). We apply a dropout rate of 0.25 to the embedding layer, hidden layers, and attention weights.

\end{document}